\documentclass[conference]{IEEEtran}
\IEEEoverridecommandlockouts
\usepackage{cite}
\usepackage{amsmath,amssymb,amsfonts}
\usepackage{algorithmic}
\usepackage{graphicx}
\usepackage{textcomp}
\usepackage{xcolor}
\usepackage{multirow}
\usepackage[super]{nth}
\def\BibTeX{{\rm B\kern-.05em{\sc i\kern-.025em b}\kern-.08em
    T\kern-.1667em\lower.7ex\hbox{E}\kern-.125emX}}
\usepackage[ruled,vlined, linesnumbered]{algorithm2e}

\begin{document}

\title{Decoding EEG--based Workload Levels Using Spatio--temporal Features Under Flight Environment
\footnote{{\thanks{This research was supported by the Challengeable Future Defense Technology Research and Development Program (912911601) of Agency for Defense Development in 2020 and was partly supported by the Institute of Information \& Communications Technology Planning \& Evaluation (IITP) grant, funded by the Korea government (MSIT) (No. 2019-0-00079, Artificial Intelligence Graduate School Program (Korea University)).}
}}
}

\author{
\IEEEauthorblockN{Dae-Hyeok Lee}
\IEEEauthorblockA{\textit{Dept. of Brain and Cognitive Engineering} \\
\textit{Korea University} \\ 
Seoul, Republic of Korea \\
lee\_dh@korea.ac.kr}

\\

\IEEEauthorblockN{Si-Hyun Kim}
\IEEEauthorblockA{\textit{Dept. of Artificial Intelligence} \\
\textit{Korea University} \\
Seoul, Republic of Korea \\
kim\_sh@korea.ac.kr}

\and

\IEEEauthorblockN{Sung-Jin Kim}
\IEEEauthorblockA{\textit{Dept. of Artificial Intelligence} \\
\textit{Korea University} \\
Seoul, Republic of Korea \\
s\_j\_kim@korea.ac.kr}

\\

\IEEEauthorblockN{Seong-Whan Lee}
\IEEEauthorblockA{\textit{Dept. of Artificial Intelligence} \\
\textit{Korea University} \\
Seoul, Republic of Korea \\
sw.lee@korea.ac.kr}
}

\maketitle

\begin{abstract}
The detection of pilots' mental states is important due to the potential for their abnormal mental states to result in catastrophic accidents. This study introduces the feasibility of employing deep learning techniques to classify different workload levels, specifically normal state, low workload, and high workload. To the best of our knowledge, this study is the first attempt to classify workload levels of pilots. Our approach involves the hybrid deep neural network that consists of five convolutional blocks and one long short--term memory block to extract the significant features from electroencephalography signals. Ten pilots participated in the experiment, which was conducted within the simulated flight environment. In contrast to four conventional models, our proposed model achieved a superior grand--average accuracy of 0.8613 ($\pm$0.0278), surpassing other conventional models by at least 0.0597 in classifying workload levels across all participants. Our model not only successfully classified workload levels but also provided valuable feedback to the participants. Hence, we anticipate that our study will make the significant contributions to the advancement of autonomous flight and driving leveraging artificial intelligence technology in the future.
\end{abstract}

\begin{small}
\textbf{\textit{Keywords--brain--computer interface, electroencephalogram, abnormal mental states, flight environment;}}\\
\end{small}

\section{INTRODUCTION}
Brain--computer interface (BCI) enables the communication between humans and devices by reflecting users’ status and intention \cite{chen2021multiattention, kim2019subject, mane2021fbcnet, reddy2021eeg, bang2021spatio, suk2014predicting}. BCI is largely divided into invasive BCI and non--invasive BCI. Invasive BCI enables to recording of brain activity via implanted electrodes close to the target neurons \cite{zhao2023modulating}. It has the advantage of representing a high signal--to--noise ratio (SNR) because it directly records brain activity, but has the disadvantage of requiring surgical operation. In contrast, non--invasive BCI enables the collection of information on brain activity without surgical operation. Compared to invasive BCI, it has the advantage of not requiring surgical operation and cost, but has the disadvantage of showing a relatively low SNR \cite{kim2015abstract, lee2019possible, lim2020unified, lee2020neural, lee2019comparative, wu2019pilots, lee2019towards, borghini2014measuring}. Thus, non--invasive BCI has been applied in various domains, including controlling external devices such as a drone \cite{lee2021design, kim2022eeggram}, a robotic arm \cite{jeong2020brain}, a wheelchair \cite{kim2018commanding}, and a speller \cite{lee2018high} and diagnosis of Alzheimer’s disease \cite{thung2018conversion}. 

In the BCI domain, detection of humans' abnormal mental states with notable performances is one of the essential issues \cite{lee2023autonomous, wu2019pilots, lee2020continuous, lim2020unified, do2019neural}. Recently, many technologies related to the technology of autonomous flight or driving have been developed for application in the real--world environment. Since the pilots’ or drivers’ mental state is one of the critical factors for the safety of passengers, the technology of accurately detecting their mental states is essential for increasing the safety. Operation of the flight is a challenging task due to the consumption of a number of energy \cite{yen2009investigation}. A positive correlation exists between workload level and task difficulty. The division of the National Aeronautics and Space Administration for Human Factors reported research indicating that as the pilots' workload increases, the ability to deal with flight--based information is decreased \cite{kantowitz2017human}.

In the case of detecting humans’ mental states, electroencephalography (EEG) signals are informative signals because they reflect humans’ status and intentions directly \cite{jeong2019classification}. Some researchers have studied detecting workload using only EEG signals. Zhang \textit{et al}. \cite{zhang2018learning} proposed the novel model that consists of deep recurrent and 3--D convolutional neural networks (CNNs) to learn EEG features across different tasks for recognizing workload. Their proposed model showed an average accuracy of 88.9 \%. Zheng \textit{et al}. \cite{zheng2023inter} proposed a cascade ensemble of multilayer autoencoders to solve the individual differences within the EEG features. Their proposed model represented a notable accuracy and computational complexity compared to various conventional methods.

The contributions of our study are as follows. \textit{i}) We designed the experimental paradigm for acquiring workload--based EEG signals under the flight environment with pilots. We induced a normal state (NS), low workload (LW), and high workload (HW) according to the difference in tasks effectively. \textit{ii}) We proposed the novel model that consists of five convolutional blocks and one long short--term memory (LSTM) block for classifying workload levels. Our proposed model showed the highest accuracy compared to the conventional models. To the best of our knowledge, this is the first attempt to classify workload levels robustly by applying the deep learning architecture.

The rest of this paper is organized as follows. In Section II, we introduce our experimental design and explain our proposed model. We present the experimental results in Section III. Finally, in Section IV, we conclude this paper.\\

\begin{figure}[t!]
\centering
\scriptsize
\centerline{\includegraphics[width=\columnwidth, height=0.30\textwidth]{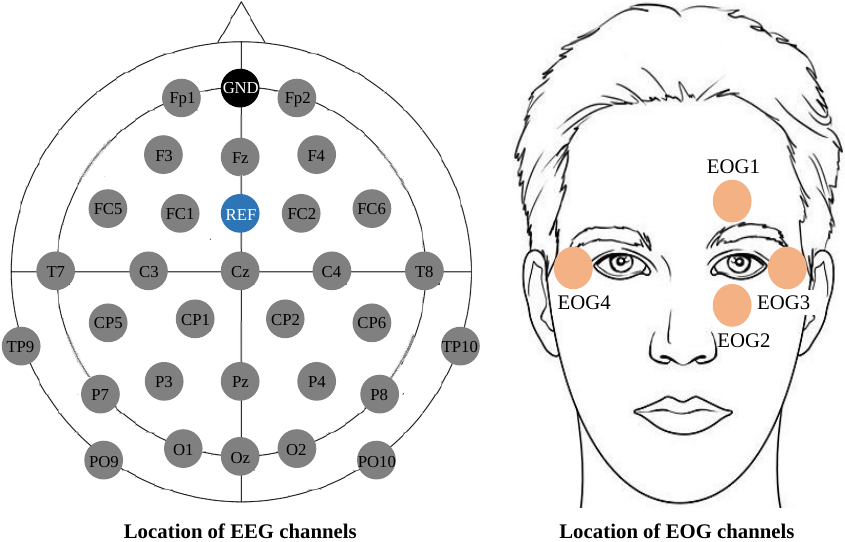}}
\caption{Data configuration of EEG and EOG channels.}
\end{figure}

\section{MATERIALS AND METHODS}
\subsection{Participants}
We acquired workload-based EEG signals from ten pilots (P1--P10, aged 25.6 ($\pm$0.52)). The criterion for recruitment was more than 100 hr. of flight experience, and they had flight experience in the Taean Flight Education Center. Also, they had no history of neurological and psychiatric disorders. Before the experiment, we informed the experimental protocols to participants, and we checked whether they had understood the experimental protocols. Additionally, they consented according to the Declaration of Helsinki. We instructed the participants to fill out a questionnaire to check their mental and physical conditions before starting and after finishing the experiment for evaluating the experimental paradigm. The Institutional Review Board of Korea University approved our experimental protocols [1040548--KU--IRB--18--92--A--2].\\

\subsection{Experimental Environment}
We used the Cessna 172 (Garmin, Olathe, KS) as the flight in the simulator which included the screen, the cockpit, and the signal amplifier. The cockpit consists of a flight yoke and other control panels for constructing the realistic flight environment. We used the signal amplifier (BrainAmp, Brain Products GmBH, Germany) for measuring EEG and EOG signals. The sampling frequency of EEG and EOG signals was set as 1,000 Hz, and a 60 Hz notch filter was used for removing direct current noise. As shown in Fig. 1, 30 EEG channels were placed according to the international 10/20 system and four EOG channels were located on the vertical and horizontal lines around the eye. AFz and FCz channels were used to the ground and reference electrodes, respectively. Before starting the experiment, we set up the impedance of all electrodes below 10 k$\Omega$ by injecting the conductive gel into the scalp of the participants.\\

\subsection{Experimental Protocol and Paradigm}
We designed the experimental paradigm to induce the pilots' workload with EEG signals effectively, as shown in Fig. 2. The experimental paradigm was designed delicately and the experiment was conducted in a strict environment to acquire EEG signals of workload. No extra time was provided to practice the tasks, since if the participants become accustomed to the task given in the experiment, the correct abnormal mental state may not occur.

We set an altitude of 3,000 feet, heading of $0\,^{\circ}$, and velocity of 100 knots as the predefined conditions of aircraft in the experiment. The instructor who sat at the next seat to the pilots gave instructions to the participants to induce the change in the workload. The instructions consisted of three levels based on the difficulty of the given task as follows.\\

\noindent
-- level 1: flight maintaining a given altitude and disregard for heading and velocity\\
-- level 2: flight maintaining continuously given altitude, heading, and velocity\\
-- level 3: steep turn with bank angle and flight maintaining velocity toward the given heading\\

The number of trials and the length of the task execution in level 1 were 15 and 60 sec., respectively. In the case of level 2, the number of trials and the length of the task execution were 10 and 60 sec., respectively. The number of trials and the length of the task execution in level 3 were 10 and 60 sec., respectively. Also, the length of the instruction was 10, 20, and 30 sec. for level 1, level 2, and level 3, respectively. The length of rest after task execution was 10 sec. regardless of the level. The instructor marked the pass or fail for the task. We defined level 1 as LW and level 2 and level 3 as HW.\\

\begin{figure}[t!]
\centering
\scriptsize
\centerline{\includegraphics[width=\columnwidth, height=0.10\textwidth]{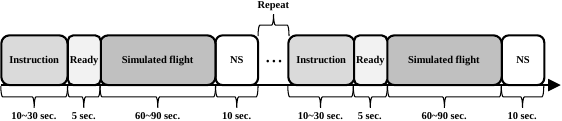}}
\caption{Experimental paradigm for inducing workload in the simulated flight environment.}
\end{figure}

\subsection{Signal Preprocessing}
We utilized a BBCI toolbox in MATLAB 2019a \cite{blankertz2010berlin} for preprocessing EEG signals. The band--pass filter was set between 1 and 50 Hz, and we conducted down--sampling of the signals from 1,000 to 100 Hz. The independent component analysis \cite{jung1998extended} was applied to remove the contaminated components of EEG signals. We segmented each trial into 1 sec. without overlap \cite{jeong2019classification}. Thus, we obtained 2,450 samples ([Level 1] 60 samples$\times$15 trials + [Level 2] 60 samples$\times$10 trials + [Level 3] 60 samples$\times$10 trials + [Rest] 10 samples$\times$35 trials) for each participant. We obtained 24,500 samples from all participants. To solve the problem that the number of samples for each state is different, we set the number of samples in LW and HW to the number of samples in NS, which has the smallest number of samples.\\

\subsection{The Proposed Model}
We proposed the deep learning--based model for classifying workload levels using EEG signals with a robust accuracy. Various features including spectral, spatial, and temporal information exist in EEG signals. Hence, we applied the principle of the hybrid deep learning framework, and our proposed model consists of five convolutional blocks and one LSTM block.

The spatio--temporal CNN in our proposed model was composed of five convolutional blocks for extracting the spatial and spectral features from EEG signals. The components of the $1^{st}$, the $2^{nd}$, and the $3^{rd}$ convolutional blocks are two convolutional layers, a 1$\times$5 filter, 1$\times$1 stride, and a batch normalization layer with a batch size of 32, 64, 128, respectively. 1$\times$5 filter is to extract the temporal information. Also, the components of the $4^{th}$ and the $5^{th}$ convolutional blocks are three convolutional layers, 5$\times$1 and 3$\times$1 filters, respectively, 1$\times$1 stride, and a batch normalization layer with the number of feature maps of 128 and 256, respectively. 5$\times$1 and 3$\times$1 filters are to extract the spatial features. To avoid the overfitting problem, we applied the maximum-- and average--pooling layers. Also, in the $5^{th}$ convolutional block, the exponential linear unit (ELU) was used as an activation function.

The data that passes through the $5^{th}$ convolutional block is input to the LSTM block. LSTM network is one of the recurrent neural networks, and it is known as the significant network for recognizing mental states. In our proposed model, one LSTM block exists, and it has two LSTM layers with 256 and 128 hidden units, respectively. We utilized the LSTM block for extracting the significant temporal features from EEG signals.

The classification block is the last part of our proposed model, and it consists of three fully connected layers and a softmax layer. The $1^{st}$ and the $2^{nd}$ fully connected layers have the hidden units of 128 and 64, respectively. Also, the output of the $3^{rd}$ fully connected layer was fed to 3--way softmax which produced a class label distribution over 3--class (NS, LW, and HW). The overall training procedure of the proposed model is summarized in Algorithm 1.\\

\let\oldnl\nl 
\newcommand{\nonl}{\renewcommand{\nl}{\let\nl\oldnl}}
\SetKwInput{KwInput}{Input}              
\SetKwInput{KwOutput}{Output}
\SetKwInput{KwStep}{Step 1} \SetKwInput{KwStepp}{Step 2}

\begin{algorithm}[!t]
\linespread{1.3}
\footnotesize
\caption{Training procedure of the proposed model}
\SetAlgoLined

\nonl $\bullet$ \KwIn{Preprocessed EEG data} 
\nonl \ \ $X$ = \{${x}_{i}\}_{i=1}^{D}$, \{${x}_{i}\}\in\mathbb{R}^{C\times T}$: Training data for mental states, where $D$ is total number of trials with $C$ the number of channels and $T$ is time points \\
\nonl \ \ $\Omega$ = \{${O}_{i}\}_{i=1}^{D}$: class labels, where ${O}_{i} \in \{NS, LW, HW\}$ and $D$ is total number of trials \\

\nonl $\bullet$ \KwOutput{Trained model}

\nonl $\bullet$ \KwStep{Training the model for NS, LW, and HW}
\ \ Input $X_{bin}$: a training EEG data\\
\ \ Input $\Omega$ = \{${O}_{tr}\}_{tr=1}^{D}$: multi--class labels, where ${O}_{tr} \in \{NS, LW, HW\}$, $D$ is total number of trials\\
\ \ The parameters of model are initialized to random values and modify the class labels to multi--class values (NS, LW, and HW) \\
\ \ Store feature maps extracted in the $5^{th}$ convolutional block\\
\ \ The loss value is generated by LSTM \\
\ \ Output $X_{N}$: network weights and loss values (multi--class) \\

\nonl $\bullet$ \KwStepp{Fine--tune parameters}
\ \ Minimizing the loss values by tuning parameters of multi--class models \\
\end{algorithm}

\section{RESULTS AND DISCUSSION}
\subsection{Performance Evaluation}
The five--fold cross--validation method was applied to evaluate an accuracy fairly. We randomly shuffled the dataset and divided it into five parts. We used four parts as the training set and one part was used as the validation set. Also, we repeated the five--fold cross--validation four times after adopting a different shuffle order each time \cite{Berrar2019cross}. We used the power spectral density--support vector machine (PSD--SVM) \cite{zhang2017design}, the DeepConvNet \cite{schirrmeister2017deep}, the EEGNet \cite{lawhern2018eegnet}, and the multiple feature block--based CNN (MFB--CNN) \cite{lee2020continuous} as the comparative models. The PSD--SVM \cite{zhang2017design} uses the PSD of the \textit{$\delta$}-- (1--4 Hz), the \textit{$\theta$}-- (4--8 Hz), the \textit{$\alpha$}-- (8--13 Hz), and the \textit{$\beta$}--bands (13--30 Hz) as a feature and the SVM as a classifier. The DeepConvNet \cite{schirrmeister2017deep} consists of four convolutional blocks, each of which is composed of a convolutional layer, a batch normalization, and an ELU activation. The $1^{st}$ block serves the dual purpose of extracting spatial and temporal features, while the other blocks are solely dedicated to extracting temporal features. The EEGNet \cite{lawhern2018eegnet} utilizes the depth--wise separable convolution, resulting in a compact parameter configuration. It is structured around two blocks, each encompassing two convolutional layers with a batch normalization, a dropout, and an ELU activation. The MFB--CNN \cite{lee2020continuous} employs deep spatial and temporal filters. To assess the influence of temporal features on an accuarcy, we specifically chose the MFB--CNN as one of the comparative models. Our model demonstrated a superior average accuracy of 0.8613 ($\pm$0.0278), surpassing an accuracy of the conventional models. Among the participants, P6 exhibited the most remarkable accuracy of 0.9214, while P10 demonstrated the lowest accuracy of 0.8251. Furthermore, our proposed model represented an exceptional level of stability, characterized by the lowest standard deviation of 0.0278, compared to the conventional models. This indicates that robustness and consistency across all participants.

\begin{table}[t!]
\centering
\caption{Comparison of an accuracy for classifying workload levels with the statistical analysis among the conventional models and the proposed model.}
\scriptsize
\renewcommand{\arraystretch}{2.0}
\resizebox{\columnwidth}{!}{
\begin{tabular}{cccccc}
\hline
Participant & PSD-SVM \cite{zhang2017design}        & DeepConvNet \cite{schirrmeister2017deep}    & EEGNet \cite{lawhern2018eegnet}         & MFB-CNN \cite{lee2020continuous}        & Proposed \\ \hline
P1          & 0.6886          & 0.7281          & 0.7332          & 0.7641          & 0.8329   \\
P2          & 0.6912          & 0.7303          & 0.7488          & 0.7732          & 0.8657   \\
P3          & 0.6204          & 0.7653          & 0.7716          & 0.7972          & 0.8604   \\
P4          & 0.7023          & 0.7602          & 0.7705          & 0.7951          & 0.8521   \\
P5          & 0.7427          & 0.8006          & 0.8214          & 0.8629          & 0.8611   \\
P6          & 0.7226          & 0.8118          & 0.8233          & 0.8602          & 0.9214   \\
P7          & 0.5897          & 0.7111          & 0.7307          & 0.7652          & 0.8438   \\
P8          & 0.7285          & 0.7882          & 0.7835          & 0.8129          & 0.8722   \\
P9          & 0.7127          & 0.7695          & 0.7701          & 0.8024          & 0.8816   \\
P10         & 0.6698          & 0.7446          & 0.7588          & 0.7831          & 0.8215   \\ \hline
Avg.        & 0.6869          & 0.7610          & 0.7712          & 0.8016          & 0.8613   \\
Std.        & 0.0485          & 0.0329          & 0.0318          & 0.0354          & 0.0278   \\ \hline
\textit{p}-value     & \textless{}0.05 & \textless{}0.05 & \textless{}0.05 & \textless{}0.05 & -        \\ \hline
\end{tabular}}
\end{table}

\subsection{Neurophysiological Analysis from EEG Signals}
We partitioned the brain region into three parts including temporal, central, and parietal regions, while categorizing the frequency into the $\delta$--, the $\theta$--, the $\alpha$--, and the $\beta$--bands. Scalp topographies were generated using EEG signals from the representative participant, denoted as P6. The grand--average band power was employed to create these scalp topographies. Both the entire array of EEG channels and each specific frequency band contributed to the calculation of amplitude. The significant variations in amplitude were observed across each brain region and spectral band. As the workload level increased (labeled as "HW"), the amplitude of the $\theta$--band exhibited an increase in the frontal region, while the $\alpha$--band amplitude decreased in the occipito--parietal region. Notably, the regions marked with the grey asterisk ($\ast$) indicate channels where the statistical significance (\textit{p}\textless{}0.05) was identified. Nevertheless, no distinctive spatial patterns were discerned in the $\delta$-- and the $\beta$--bands.

\begin{figure}[t!]
\centering
\scriptsize
\centerline{\includegraphics[width=\columnwidth, height=0.40\textwidth]{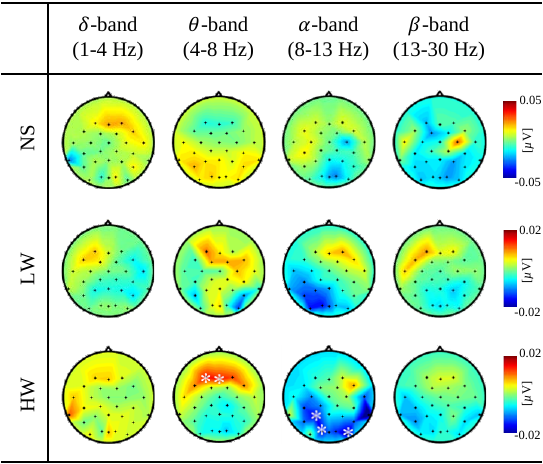}}
\caption{Scalp topographies according to the spectral bands (\textit{$\delta$}--, \textit{$\theta$}--, \textit{$\alpha$}--, and \textit{$\beta$}--bands) across the representative participant (P6). The locations of channels with the statistical significance are indicated as grey `$\ast$' (*: \textit{p}$<$0.05).}
\end{figure}

\section{CONCLUSION}
Among the various factors causing aviation accidents, the mental state of pilots is a critical factor. Given its direct implications for the safety of passengers, the imperative lies in achieving a robust detection of pilots' mental states. Detection of various mental states through EEG signals is one of the important issues in the BCI domain. Notably, a nuanced classification of mental states is important to prevent accidents caused by humans. This study introduces the novel model designed for the classification of workload levels using EEG signals. Our proposed model demonstrates a superior accuracy in classifying workload levels, compared to the conventional models. With the acquisition of workload--based EEG signals from pilots, our research will contribute to the development of technology in autonomous flight and driving. We plan to develop our proposed model through the acquisition of EEG signals associated with various abnormal mental states using our modified experimental protocols to apply our research to the real--world environment.\\

\bibliographystyle{IEEEtran}
\bibliography{REFERENCE}

\end{document}